\documentstyle[11pt]{article}
  \advance\oddsidemargin  by -1in
  \advance\evensidemargin by -1in
  \advance\topmargin	  by -0.3in
\def\lsim{\mathrel{\rlap{\lower4pt\hbox{\hskip1pt$\sim$}}
    \raise1pt\hbox{$<$}}}	  
\def\gsim{\mathrel{\rlap{\lower4pt\hbox{\hskip1pt$\sim$}}
    \raise1pt\hbox{$>$}}}	  

\def\beq{\begin{equation}}
\def\eeq{\end{equation}}
\def\arr{\begin{eqnarray}}
\def\endarr{\end{eqnarray}}

\makeindex
\begin{document}
\hspace{12cm}
{\large LPTHE 93-16}
\vspace{2cm}\\
\begin{center}
{\Large \bf Hadronization in Nuclear Matter}
\vspace{0.5cm}\\
{\large Boris Kopeliovich\\
 \it Laboratoire de Physique Th\'eorique et Hautes \'Energies\\
\it Universit\'e de Paris Sud, Orsay, France\\
and\\
\it Joint Institute for Nuclear Research, Dubna, Russia\\
\it e-mail: boris@bethe.npl.washington.edu}\\
\vspace{2.5cm}
{\large \bf Abstract}
\end{center}
\date{ }
Nuclei are unique analyzers of the space-time development of jets at
early stage.  We argue that the gluon bremsstrahlung, rather than the
color string, is the main mechanism of hadronization of highly virtual
quarks produced in a hard interaction.	It results in an energy- and
time-independent density of energy loss, like a color string, but
steeply dependent on the quark virtuality.  Effects of formation zone
(FZ) and color transparency (CT) substantially affect the jet
quenching in a nuclear matter.	The latter also plays an important
role in the broadening of transverse momentum distribution of a quark
passing a nucleus.  Parameter-free calculations provide a good
description of available data on nuclear effects in the leading hadron
production in deep-inelastic lepton scattering, back-to-back
high-$p_T$ hadron pair production, broadening of the transverse
momentum distribution in the Drell-Yan process of lepton pair
production on nuclei.
\vspace{4cm}\\
\begin{center}
{\sl Talk presented at the Workshop\\
Future Directions in Particle and Nuclear Physics\\ with Multi-GeV
Hadronic Beam Facilities\\
Brookhaven, March 4-6, 1993}\\
\end{center}
\newpage
\section{Introduction}
Hadronization of highly virtual quarks is studied in
$e^+e^-$ annihilation, deep-inelastic lepton scattering,
high-$p_T$ jet production in hadronic interaction.  Assuming
factorization one can describe the hadronization stage
phenomenologically by a fragmentation function $D(x,p_T)$ of
a quark, which is a distribution of produced hadrons over
share of the quark momentum, $x$, and transverse momentum,
$p_T$.

\subsection{Why nuclear target?}

A quark knocked out of a parent hadron in a hard scattering,
converts into colorless hadrons in a while due to the phenomenon of
confinement.  The Lorentz time delay considerably stretches the
duration of this process, proportionally to the initial quark
momentum.  Hadrons created in the final state carry poor information
about dynamics of hadronization.  Most important details are hidden at
its early stage.  A nuclear target provides a unique opportunity to
look inside the process at microscopic times after it starts.  The
quark-gluon system, originated from a quark knocked out in a hard
collision, interacts while passing a nucleus.  This can bring
forth precious information about the structure of this system and its
space-time development.
On the other hand, the hadronization could be a sensitive probe of the
state of the nuclear matter.

\subsection{What energies?}

The fundamental principle closely related to hadronization is the
Landau-Pomeranchuk phenomenon \cite{L-P}: multiple interaction of a
projectile
particle does not affect the emission of long-waved photons
(gluons). In other words, a nucleus does not disturb a spectrum of
gluons, which have time of emission much longer than the nuclear
radius.
\begin{equation}
t\approx\frac{\omega}{k_T^2}\gg R_A ,
\label{aa}
\end{equation}
Here $\omega$ and $k_T$ are the energy and transverse momentum of a
gluon. Note that it concerns only hard reactions, in soft interactions
a nuclear target affects the particle production due to so called
nonplanar diagrams \cite{Cap}.

It follows from (\ref{aa}) that in order to study the nuclear effects
in hadronization one does not need superhigh energies, it is
sufficient to have
\begin{equation}
E\leq {k_T^2}\;R_A ,
\label{ab}
\end{equation}
what is $50-100\;GeV$ if $k_T^2\approx 1\;GeV^2$.

\subsection{What reactions?}

Probably the most direct way to study the nuclear effects in
hadronization has a process of jet production in deep-inelastic lepton
scattering on a nucleus. Measurement of lepton momenta provides
information about virtuality and energy of a quark initiated the jet.

High-$p_T$ hadron production in hadron-nucleus and nuclei collisions
provides additional information about the nuclear modification of
structure function of an incoming hadron. Multiple interactions are
essential in the inclusive hadron production with high $p_T$.

An effective way to suppress multiple rescattering effects is the
study of production of symmetric hadron pairs with high-$p_T$ on
nuclei.

Multiple interaction of a quark propagating through nuclear matter
leads to the increase of its transverse momentum. An undisturbed
information about it brings forth the study of the nuclear broadening
of transverse-momentum distribution of Drell-Yan lepton pairs.

Nuclear shadowing and the broadening of transverse momenta of heavy
quarkonia also can provide a precious information about hadronization
in a nuclear matted.

\subsection{What are the observables of hadronization?}

Usually experimental data are
represented in the form of ratio of nucleus to nucleon cross sections,

\begin{equation}
R_A=\frac{\sigma^A(x,p_T)}{A\;\sigma^N(x,p_T)} ,
\label{ac}
\end{equation}
where $x$ is the Feynman variable relative the initial quark
momentum in the case of DIS, or incoming hadron.
This simple quantity, $R_A$, nevertheless contains rich
information about dynamics of hadronization.

\section{Gluon bremsstrahlung versus string model}

\subsection{Energy loss in a nuclear matter}

{\sf Hard gluon bremsstrahlung}.  The hard reaction with a large
square
of momentum transfer $Q^2$ cannot resolve the quark structure at small
impact parameters, $b^2<Q^2$, and knocks out the quark together with a
hard transverse components of its color field, $k_T^2>Q^2$.  However
the softer part of the quark color field, at impact parameters
$b^2>Q^2$ is "shaken off" in the form of gluon bremsstrahlung.
  This qualitative consideration
demonstrates that there should be a steep $Q^2$-dependence of the
retarding force due to the hard gluon bremsstrahlung.
It was demonstrated in \cite{Nied,KN-JINR,KN-old,KN-new},
that the bremsstrahlung like a string produces a constant density of
energy loss

\begin{equation}
\kappa_{br}=-\frac{dE}{dz}=\frac{2}{3\pi}\,\alpha_s[Q^2(t)]\,Q^2(t)
\label{ad}
\end{equation}

Note that the time-independent energy loss (\ref{ad}) is the
result
of the high-energy approximation.
Corrections important at moderate energies were introduced in
\cite{KN-old,KN-new}.

{\sf String model}.  It is assumed that even a highly virtual
quark, knocked out in a hard process forms a stationary color tube
\cite{CNN,KNied,Kop-rev,Bialas,Gyulassy'}.  Properties of a string
stretched between a $q$
and $\bar q$ (diquark), flying apart, are the same as for the static
system.  This assumption fixes the density of energy loss,
$dE_q/dz=-\kappa\approx 1\,GeV/fm$.  It is assumed to be independent
on the quark virtuality, $Q^2$.

It is difficult to justify these assumptions and the ignorance of the gluon
bremsstrahlung, especially at high $Q^2$.

\subsection{Formation zone of leading hadron production}

During hadronization a quark looses energy for hadron production,
until its color is neutralized.  The produced colorless wave packet
develops a hadron wave function after a while.	We call hereafter the
time of color screening, the formation zone (FZ). It plays an
important
role in nuclear attenuation of leading hadron.	Indeed, an inelastic
interaction of the produced colorless wave packet induces new energy
loss and a strong attenuation, rather than reinteraction of the quark
before the color neutralization \cite{Kop-PL}.

The crucial point is the behavior of the FZ, $l_f$, at
$x\rightarrow 1$.
Energy conservation imposes the restriction,

\begin{equation}
l_f \leq\frac{\nu}{\kappa} \;(1-x)
\label{ae}
\end{equation}

Indeed, the quark radiates hadrons and looses the energy during
FZ, hence at $x_F\rightarrow 1$ it has to convert into a
colorless state shortly.  This restriction was found in \cite{KL,KNied}
and confirmed in \cite{BG} using
Monte-Carlo simulation of the string decay.

Expression (\ref{ae}) has general character.  The models under
discussion, differ only
in a value of energy loss density, $\kappa$.
Since the energy loss of a highly virtual quark may
be much higher than $1\;GeV/fm$, due to the gluon bremsstrahlung,
FZ is substantially shorter than is expected in the
string model.

\subsection{Attenuation of a quark in nuclear matter}
A high-energy quark cannot
be absorbed, because soft color-exchange rescatterings provide no
longitudinal momentum transfer.  However each color-exchange
scattering of the quark initiates an additional production of
low-energy particles, what leads to an increase of the energy loss,
i.e.  attenuation.  It was found in \cite{Kop-PL} that due to this
effect a
nuclear matter only slightly modifies the quark fragmentation function
at high energy and long FZ, $l_f\gg R_A$, but produces a
dramatic effect at intermediate energies.

\subsection{ Nuclear attenuation of a colorless wave packet}
At this point two models under consideration differ crucially.
The colorless wave packet attenuates with an inelastic cross
sections,
which depends on its transverse size due to color transparency (CT)
\cite{ZKL}. At
$x\rightarrow 1$, a quark of virtuality $Q^2$, produces according to
(\ref{ae}), a colorless wave packet instantaneously.  So it has a
small transverse size, about $1/Q^2$, and does not attenuate.
In the case of a finite
FZ the quark gradually looses its virtuality, and
produces a wave packet of virtuality \cite{KN-old,KN-new},

\begin{equation}
Q^2(l_f)=\frac{Q^2\,\kappa}{\kappa+Q^2\,(1-x)}
\label{af}
\end{equation}
It follows from (\ref{ad}) and (\ref{af}) that the initial size of the
colorless wave packet depends mainly on $x_F$.

To the contrary, in the naive version of {\it string model} considered
here, the CT effects are neglected, and it is assumed
that the produced colorless wave packet attenuates with the hadronic
cross section, independently on $Q^2$ \cite{Bialas,Gyulassy'}.  Note
that in more realistic
version of string model the transverse size of the produced state is
also small at $x\rightarrow 1$, similar to (\ref{af}), because the
distance between the leading quark and the kink, produced in the color
field of the target vanishes proportionally to $(1-x)$ (X.Artru,
private communication)

\section{Deep-inelastic scattering}

In this case the ratio $R_A$, (\ref{ac}), is simply the ratio of the
quark fragmentation function in nucleus to the one in vacuum.
Some corrections come from the modification of the
structure function of bound nucleon (EMC-effect).  We compare in
fig.~1a the parameter-free calculations \cite{KN-old,KN-new} of
energy-dependence
of $R_{Cu}$ with the data \cite{EMC,SLAC} averaged over interval
$x>0.2$.
$x$-dependence is presented in fig.~1b, as compared with high energy
data \cite{EMC}. Note
that
the naive version of string model \cite{Bialas,Gyulassy'},
predicts a strong
decrease of $R_A$ towards $x=1$, in the region which is not covered
by the available data.	The $Q^2$-dependence of $R_A$ is compared with
data \cite{EMC} in fig.~1c.  Curves calculated with zero FZ,
or neglecting CT, demonstrate the relative role of
these phenomena.  One can see that CT is more
important for the observed suppression of nuclear shadowing.
Actually, the $Q^2$-dependence is flat, just due to CT.  The decrease
of $R_A$ at $Q^2>40\;GeV^2$ comes from the EMC effect in the nuclear
structure function.

\section{High-$p_T$ hadron production}

{\sf High-$p_T$ probe of quark-gluon plasma}. I was argued in
\cite{Gyulassy}
that energy loss of a quark propagating through a nuclear matter is
sensitive to the temperature of the medium. At high temperature a
quark-gluon plasma may be formed and integration over $k_T$ of gluon
bremsstrahlung is cut by the reversed Debye screening radius, $\mu$.
In this additional energy loss induced by the medium,
$\Delta\kappa\propto\mu^2$ is sensitive to the temperature. However in
inclusive production of particles with high $p_T$ the quark multiple
rescattering corrections are out of a control. They are just the
reason of the "Cronin effect", nuclear antishadowing at high $p_T$.
So the large momentum transfer, $p_T\gg\mu$, is shared between all
rescattering and the sensitivity to the value of $\mu$ vanishes.

{\sf Symmetric pair production}.  As was mentioned, symmetric
production of particles with high $p_T$ on nuclei allows to suppress
the influence of multiple rescatterings.  However no sensitivity to
the Debye screening appears in this case.  The integration over $k_T$
is restricted in this case by an acceptance of a spectrometer, which
defines the nuclear quenching of the symmetric jets.

It is argued in \cite{KNied,KK} that at
moderate values of
$p_T<1-2\;GeV$ the main contribution to back-to-back particle
production comes
from an uncorrelated production of two particles, in symmetric
configuration.	This follows from $p_T$-dependences of the correlation
function and the slope.  Using data on $A$-dependence of inclusive
production of hadron with high $p_T$, one can calculate this
uncorrelated contribution to the cross section of pair production
without free parameters.  The results are presented in fig.~2 for the
exponent of $A^{\alpha_2}$-dependence of the cross section (a-c) and
the ratio $R_{W/Be}$.  This contribution is responsible for the bumps
observed
at moderate values of $p_T$. The data \cite{Abr,McC,Fin,Hsi,Str,Straub}
confirm the existence of this contribution at all available energies.
The contribution of the hard, back-to
back scattering is shown in fig.2a by the dashed curve \cite{KN-pT}.
It is
a parameter-free calculation, analogous to the previous case of DIS.
The main difference is a smearing of the initial quark momenta,
weighted with the hadron structure functions.  Another distinction is
energy loss of the projectile quark participating in the hard
scattering in the initial state.  It is reduced to the nuclear
modification of the projectile hadron structure function, and
also induces an attenuation.

\section{Nuclear broadening of the transverse-momentum distribution}

A quark propagating through a nuclear matter increases its transverse
momentum due to the multiple rescattering on bound nucleons.  An
important phenomenon affecting this process is CT.
Due to confinement the color of a quark is always compensated by the
colors of other accompanying partons.	While this fact is unimportant
at high momentum transfers, it is crucial for soft processes.
It cuts off all soft gluons whose wavelength
is longer than the color screening-radius \cite{DHK}.
The remarkable conclusion of \cite{DHK} is the universality
of the broadening of $<p_T^2>$ which is independent on the radius of
color screening, $r$. It is the direct consequence of the color
screening: the smaller is $r$ the rarer are the quark rescatterings,
but the larger is $\Delta <p_T^2>$ in each interaction.

{\sf Drell-Yan process}. The momentum distribution
of a quark after the multiple interaction with the target
nucleons can be measured in a Drell-Yan process of lepton-antilepton
pair production.  Since the lepton pair does not interact on its way
out of the nucleus, it carries the undistorted information about the
interactions of the quark.
The experiment NA10 \cite{NA10} has measured a value
$\Delta<p^2_T>^{\pi W}_{DY}=0.15\pm.03\mbox{stat}\pm.03\mbox{syst
(GeV/c)}^2 $
for incoming pions at 140 and 286~GeV, while
the experiment E772 with 800~GeV protons \cite{E772} reports a value
$\Delta<p^2_T>^{pW}_{DY}=0.113\pm.016\ \mbox{(GeV/c)}^2$.
The two values coincide within the error bars which
implies that indeed no dependence on the type of incident hadron and
its energy is visible.
The parameter-free calculation gives,
$\Delta<p^2_T>^{hW}_{DY}=0.17 \mbox{(GeV/c)}^2$
in fair agreement with the data.

The results of calculations \cite{DHK} of
the normalized ratio of differential cross
sections of the Drell-Yan
process on nucleus to nucleon targets, $R(A/N)$,
 are compared with experimental data
\cite{NA10,Adle} on fig.~3. One can see that this parameter-free
calculation also
provides a good description of the experimental data.

{\sf Deep-inelastic scattering}. The broadening of the
transverse momenta of produced hadrons, $\Delta<p^2_T>$, is
smaller by factor of $x^2$ than that of the quark. The latter is
expected to be the same at high FZ as in the Drell-Yan
process. No broadening was observed in \cite{EMC-old} at high
energies. It
obviously is connected with a small value of $<x^2>\approx 0.075$ in
this data. So we expect the usual "sea-gull" behavior of x-dependence
of $\Delta<p^2_T(x)>_h$ at high energies, with the same
$\Delta<p^2_T>_q$ as in Drell-Yan process. According
to results of \cite{DHK} no $Q^2$-dependence of
$\Delta<p^2_T(x)>_h$ is expected,
except $x\rightarrow 1$, where FZ is small, and decreasing
$Q^2$-dependence is predicted \cite{KN-new}.

\section{Conclusions}

Existing theoretical approaches to the hadronization dynamics and its
modification in a nuclear matter, though being quite rough, provide a
good description of available experimental data with a small number of
parameters, or parameter-free. Even the naive version of string
model, facing severe problems at a theoretical level, does not
contradict the data. It is a consequence of the lack of detailed and
high-statistics measurements. New experimental study of of high-$p_T$
hadron production with large $x_T$ and different flavors,
electroproduction of hadrons
with measurement of $\nu-, Q^2-, x-$ and $p_T$ distributions, on
nuclear targets are highly desirable.

{\sf Acknowledgements}. The author thanks LPTHE of the University of
Paris for the hospitality, and CNRS for the financial support during
this study.

 \end{document}